# Gate-tuned Superconductor-Insulator transition in (Li,Fe)OHFeSe

B. Lei[1], Z. J. Xiang[1], X. F. Lu[1], N. Z. Wang[1], J. R. Chang[1], C. Shang[1], X. G. Luo[1,4], T. Wu[1,4], Z. Sun[3,4] and X. H. Chen[2,1,4]

1. Hefei National Laboratory for Physical Sciences at Microscale and Department of Physics, University of Science and Technology of China, Hefei, Anhui 230026, China and Key Laboratory of Strongly-coupled Quantum Matter Physics, Chinese Academy of Sciences, Hefei, Anhui 230026, China,
2. High Magnetic Field Laboratory, Chinese Academy of Sciences, Hefei, Anhui 230031, China
3. National Synchrotron Radiation Laboratory, University of Science and Technology of China, Hefei, Anhui 230026, China
4. Collaborative Innovation Center of Advanced Microstructures, Nanjing 210093, China

**The antiferromagnetic (AFM) insulator-superconductor transition has been always a center of interest in the underlying physics of unconventional superconductors. The quantum phase transition between Mott insulator with antiferromagnetism and superconductor can be induced by doping charge carriers in high-$T_c$ cuprate superconductors[1]. For the best characterized organic superconductors of $\kappa$-(BEDT-TTF)$_2$X (X=anion), a first order transition between antiferromagnetic insulator and superconductor can be tuned by applied external pressure or chemical pressure[2-4]. Also, the superconducting state can be directly developed from antiferromagnetic insulator by application of pressure in Cs$_3$C$_{60}$[5]. The resemblance of these phase diagrams hints a universal mechanism governing the unconventional superconductivity in close proximity to antiferromagnetic insulators. However, the superconductivity in iron-based high-$T_c$ superconductors evolves from an antiferromagnetic bad metal by doping charge carriers, and no superconductor-insulator transition has been observed so far[6-8]. Here, we report a first-order transition from superconductor to insulator with a strong charge doping induced by ionic gating in the thin flakes of single crystal (Li,Fe)OHFeSe. The superconducting transition temperature ($T_c$) is**

**continuously enhanced with electron doping by ionic gating up to a maximum $T_c$ of 43 K, and a striking superconductor-insulator transition occurs just at the verge of optimal doping with highest $T_c$. A novel phase diagram of temperature-gating voltage with the superconductor-insulator transition is mapped out, indicating that the superconductor -insulator transition is a common feature for unconventional superconductivity. These results help to uncover the underlying physics of iron-based superconductivity as well as the universal mechanism of high-$T_c$ superconductivity. Our finding also suggests that the gate-controlled strong charge doping makes it possible to explore novel states of matter in a way beyond traditional methods.**

Phase diagrams have been extensively studied in the iron-based superconductors with a large range of carrier doping, while such investigations are very limited in the iron chalcogenides because the known FeSe-derived superconductors are not suitable to study the underlying physics due to reasons mentioned in the Ref.10. On the other hand, controllable manipulation of various phases through the gate-tunable charge doping has been confirmed to lead to a new device paradigm for future material sciences and technologies[11-14]. Recently, we find a ionic liquid that can dope large charge carriers with driving Li ions in the layered 1T-TaS$_2$ controlled by a gate electric field in the extreme charge-carrier-concentration limit[15], which is different from the ionic liquid used for the electron-double layer surface gating. Such ionic liquid provides us a chance to map out the phase diagram in the layered iron-chalcogenide superconductors. The newly discovered FeSe-derived superconductor (Li$_{0.8}$Fe$_{0.2}$)OHFeSe[10] is suitable for mapping out a carrier-doping phase diagram by ionic gating due to its stability in air and layered structure. In this letter, we show the variation of transport behavior with carrier concentration controlled by gate electric field in the thin flakes of (Li,Fe)OHFeSe. A remarkable superconductor-insulator transition occurs with the carrier doping, bearing resemblance to phase diagrams of many other unconventional superconductors.

We have successfully achieved the continuous gate voltage controlling of the electronic phases in the layered high-$T_c$ superconductor (Li,Fe)OHFeSe with a setup of ionic field-effect transistor (iFET). A schematic configuration of our iFET device is shown in Fig. 1a. Electrical transport properties of exfoliated single-crystalline (Li,Fe)OHFeSe thin flakes with typical thickness of ~200 nm were systematically studied upon applying a gate voltage ($V_g$). A liquid polymer electrolyte (LiClO$_4$/PEO) was chosen to serve as the dielectric through which the electric field was established. The details of device preparation are described in the Methods. As a continuously swept positive gate voltage is applied at 330 K, the resistance of (Li,Fe)OHFeSe thin flake starts to drop gradually when $V_g$ is ramped up to about 2.5 V, and reaches a minimum at $V_g \approx 4$ V, then increases rapidly and is enlarged by one order of magnitude as $V_g$ approaches 5 V as shown in Fig. 1b. This non-monotonic behavior hints that an unusual change takes place around $V_g$ =4 V, separating the low-resistive state and high-resistive one.

To investigate the nature of this unusual change, we measured the temperature dependence of resistance of (Li,Fe)OHFeSe with various $V_g$ applied on both sides of the critical gate voltage $V_g$ =4 V (see Fig. 2). At $V_g$ = 0 V, this sample is superconducting (SC) with mid-point critical temperature $T_c^{mid}$=24.4 K. The zero-resistance temperature ~20 K is consistent with magnetic susceptibility measurement (see Supplementary Section I). For a comparison, the $T_c$ of polycrystalline sample (Li$_{0.8}$Fe$_{0.2}$)OHFeSe is always ~ 43 K[10,16]. The significant lower $T_c$ in the single crystals is likely due to the vacancies of Li, non-ideal atomic occupancy and different

Li/Fe ratio in the hydroxide layer (Li,Fe)OH, which strongly depend the growth process. With the increase of $V_g$, $T_c^{mid}$ continuously shifts up and the normal-state resistance decreases until a critical gate voltage $V_g^C$ = 3.9 V. At $V_g = V_g^C$ the sample exhibits a maximum of $T_c^{mid}$= 43.4 K, being the same value as we have observed in poly-crystalline $(Li_{0.8}Fe_{0.2})OHFeSe$[10,16]. Beyond $V_g^C$ an insulating behavior immediately sets in and the SC transition completely disappears, as indicated by the R-T curve at $V_g$ = 4.0 V. Further increase of $V_g$ promptly strengthens the insulating behavior, and finally pushes the low-temperature resistance to a magnitude of more than five orders larger than the normal-state resistance of the SC phase. In all devices we studied, $T_c^{mid}$ at $V_g$ = 0 V varies from 22 K to 36 K (see Supplementary Information, Fig. S1b), and $V_g^C$ also ranges from 3.6 V to 4.0 V, but the overall evolving behavior upon gating always remains the same. Especially, both the optimal $T_c^{mid}$~ 43.4 K and the sharp transition from the SC to insulating behavior were repeated in each device. No mixed state of SC phase with optimal $T_c$ and the insulating phase has ever been observed.

Considering the sharpness of SC transition (transition width < 1 K) in the vicinity of optimal superconductivity (Supplementary Information, Fig. S3a), and the absence of residual superconductivity in weak insulating phase, we conclude that the electronic properties throughout the whole sample flake remain essentially homogeneous. Since the sample thickness is about 200 nm, much larger than the typical effective modulation depth of electrostatic double layer gating (usually only a few nm[17-20]), we can reasonably attribute the gate-controlling effect in our iFET to Li ion doping rather than electrostatic charge accumulation. There are other evidences to support the electrochemical doping mechanism—the tuning process is not fully reversible if the $V_g$ was overloaded. We can only restore the initial zero-bias resistance and superconductivity when $V_g$ was swept back before reaching a threshold voltage $V_{th} \leq V_g^C + 0.3$ V, corresponding to resistance $R \sim 35\text{-}45\ \Omega$ at 330 K. When $V_g$ exceeds $V_{th}$, the insulating phase would be preserved even if $V_g$ was swept back to zero or even -6 V. This behavior suggests that a large gate voltage $V_g > V_{th}$ can cause chemical modification of (Li,Fe)OHFeSe, inducing an irreversible transition in the insulating phase regime. Detailed discussion of the effect of applied gate voltage is presented in the Supplementary Information.

To further reveal the underlying electronic properties, we measured the temperature and gate voltage dependences of Hall resistance $R_{xy}$ as shown in Fig. 3a. The Hall coefficient $R_H$ in SC samples is negative, and decreases upon cooling, suggesting the transport properties of superconducting (Li,Fe)OHFeSe are dominated by electron-type carriers at low temperatures. At $V_g = 0$ V, $R_H$ shows a flat minimum at $\sim 75$ K then increases slightly with decreasing temperature. As $V_g$ is raised up to 3.6 V, this minimum disappears, and for the optimal gating $V_g = 3.9$V ($T_c^{mid}=$ 43.4 K), $R_H$ decreases rapidly toward $T_c$. The low-temperature downturn of $R_H$ has also been observed in $K_xFe_{2-y}Se_2$ single crystals with $T_c = 32$ K[21]. Besides, the evolution of $R_H$-$T$ curves from undoped ($V_g = 0$ V) to optimally doped ($V_g = 3.9$ V) sample mimics that of 1-uc FeSe films on SrTiO$_3$ substrate with annealing time prolonged from 20 h to 42 h[22]. This similarity may hint some universal feature in these high-$T_c$ Fe-chalcogenide superconductors. In the insulating regime, the field dependence of $R_{xy}(H)$ becomes nonlinear and dominated by hole-type carriers (see the Supplementary Information for more details), being similar to the behavior of bulk β-FeSe[23].

In Fig. 3b, we plot $R_H$ and Hall number $n_H = 1/eR_H$ as a function of $V_g$ at $T = 70$ K. We note that the Hall number per Fe site (Fe(2) site) of the conducting FeSe layer is appreciable, thus could not represent the actual doping level. Similar to many of iron-based superconductors[23,24], $R_H$ in (Li,Fe)OHFeSe could be significantly reduced owing to the compensating effect between different bands. Upon applying $V_g$, $R_H$ initially increases in accord with an electron doping process. At $V_g = 3.8$ V, where $T_c^{mid}$ increases to 38.6 K, $R_H$ reaches a maximum where the Hall number $n_H$ suggests an extra doping of ~0.1 $e$ per Fe(2) site compared to the ungated state. With further increasing $V_g$, the optimal $T_c$ can be achieved, whereas $R_H$ decreases drastically, being consistent with the low-temperature downturn of $R_H$ in the near optimally gated sample in Fig. 3a. Referring to the obvious decrease of the low-temperature resistance in this gate voltage range (see Fig.2), we attribute this feature to the enhanced electron mobility near the optimal doping. The most remarkable part in our data is the sudden sign change of $R_H$ across the boundary of the superconductor-insulator (S-I) transition, suggesting a dramatic modification in the electronics across the phase boundary.

We note that the gate-controlled S-I transition does not refer to any thickness effect. The general thickness of sample flakes used in our iFET corresponds to about 200 unit cell, which is thus far away from the two dimensional limit. All the behavior observed upon applying gate voltage reflects the properties of bulk (Li,Fe)OHFeSe samples in a gate-induced electrochemical lithiation, whose modulation depth is significantly larger than most of the existing electric-field-controlling techniques. Devices based on the similar ion insertion effects are envisaged to be promising in the precise controlling of physical parameters and elaborately studying of systems with complex phase diagrams.

Despite the difficulties of characterizing the iFET *in situ* during the gate-voltage tuning process, some findings are helpful for the understanding of the origin of the S-I transition and the nature of the gate modulation mechanism. The experimental evidence suggests a gate-controlled lithiation scenario: (i) Below $V_{th} > V_g^C$, the transition is reversible, we can drive the iFET to go through the phase boundary repeatedly (Supplementary Information), which distinguishes the S-I transition from the possible sample degradation. (ii) In the samples with irreversible insulating behavior, we could find neither detectable shifting (> 0.01°) of X-ray diffraction peaks nor new Raman modes (Supplementary Information). These results unambiguously rule out the possibility of both degradation and Li ion intercalation into the space between the hydroxide layer and FeSe layer. (iii) The SC transition width is obviously reduced as approaching the optimal $T_c$, indicating the improvement of sample homogeneity upon gate modulating in SC regime.

Based on these findings, we propose a scenario of lithiation process. First, the Li ions in the electrolyte enter (Li,Fe)OHFeSe and fill the vacancies in the (Li,Fe)OH space layer, acting as an electron donor to the FeSe layers. Both the reduction of the vacancy concentration and the electron doping can contribute to the enhancement of $T_c$. Similar controlling of $T_c$ by slightly changing the $Li^+/Fe^{2+}$ ratio in the hydroxide layer has been reported in powder samples[27,28]. As most of vacancies are filled, the samples achieve the optimal $T_c \sim 43.4$ K at $V_g = V_g^C$ and becomes almost homogeneous as indicated by the sharp superconducting transition. Between $V_g^C$ and $V_{th}$, the (Li,Fe)OHFeSe thin flakes suddenly become insulating, which is very surprising and will be discussed below. With $V_g$ continues ramping up beyond $V_{th}$, Li ions will start to replace the Fe atoms in (Li, Fe)OH. Because the "displaced out" Fe ions have a much lower mobility than Li ions in the electrolyte, they can hardly re-enter the lattice and re-occupy the original sites when $V_g$ was swept back, which eventually results in an irreversible state.

The phase diagram of gate-voltage-modulated (Li,Fe)OHFeSe iFET is depicted in Fig. 4. Two distinct phase regimes, the metallic phase with superconductivity emerging at low temperatures and the insulating phase, are separated by a sharp phase boundary located at $V_g = V_g^C \sim 4$ V without any evident coexistence of these phases. The abrupt disappearance of $T_c$ and discontinuity of $R_H$ at the phase boundary indicate a first-order character of this gate-voltage induced S-I phase transition. In analogy with other unconventional superconductors induced by doping a parent compound, the phase diagram could be interpreted by doping holes into the insulating phase that is in the gating voltage range from $V_{th}$ to $V_g^C$. In a reversible cycle, by tuning the gate voltage from $V_{th}$ to $V_g^C$ and lower value, the driven-out Li ions from the crystals can effectively dope holes into the electronic system, and eventually makes the insulating state melt down and gives rise to superconductivity. Though the exact nature of the insulating phase is still under investigation, one plausible scenario is that excess electron charges in the FeSe layers may give rise to charge localization due to some electron correlation effects.

The most intriguing feature is the appearance of an insulating phase immediately adjacent to the optimal superconductivity, which is reported for the first time in iron-based superconductors. Similar phase diagrams have been observed in the two-dimension organic superconductors[2-4] and $Cs_3C_{60}$[5]. This phase diagram of FeSe-derived superconductors also bears resemblance with that of high-$T_c$ cuprate superconductors. The similarity of these phase diagrams transcends the diversity of various unconventional superconducting materials, suggesting that all of them share a universal mechanism in superconductivity. Our finding in (Li,Fe)OHFeSe iFET helps to unify the underlying physics in both the cuprates and FeSe-derived materials. Moreover, our work suggests that the gate-controlled strong charge doping is a very powerful practice for the exploration of novel states of matter that cannot be realized using traditional methods.

## Methods

### Device fabrication

Single crystals of pristine (Li,Fe)OHFeSe were grown by the hydrothermal reaction method. 0.96 g selenourea (Alfa Aesar, 99.97% purity), 0.2094g Fe powder (Aladdin Industrial, analytical reagent purity), and 6g LiOH·$H_2$O (Sinopharm Chemical Reagent, analytical reagent purity) were put into a Teflon-lined steel autoclave (50ml) and mixed with together with 10 ml de-ionized water. The Teflon-lined autoclave was than tightly sealed. The sealed autoclave was heated to 160°C at a rate of 0.5°C/min and kept at 160°C for 3 days, then slowly cooled to room temperature with 2°C/h. Shiny flake crystals with typical size of 0.3 × 0.3 × 0.02 mm$^3$ were obtain by repeatedly washing the product with de-ionized water. To obtain ($Li_{0.8}Fe_{0.2}$)OHFeSe thin flakes, we mechanically exfoliated single crystal pieces using the scotch tape method onto a $SiO_2$ insulating layer with a thickness of 300 nm grown on a highly doped Si substrate. We chose the thin flakes with proper shape and flatness using an optical microscopy, and characterized the thickness by an atomic force microscopy. Using the lithography and lift-off techniques, thin flakes (typically 150 to 230 nm thick) were patterned into a standard Hall bar configuration and coated with Ti/Au electrodes (10/250 nm) for transport measurements.

A mixture of LiClO$_4$ and polyethylene oxide (LiClO$_4$/PEO) was used as the solid electrolyte. LiClO$_4$ (Alfa Aesar) and PEO (Alfa Aesar) powders (0.3 g and 1 g, respectively) were mixed with 15 ml anhydrous methanol (Alfa Aesar). The mixture was then stirred at 60 °C for at least 10 hours in a glove box filled with high pure argon atmosphere. When applying the electrolyte, the mixture was dropped onto the Si wafer covering both the sample and the gate electrode, then baked on a hot plate at 90°C for about 10 minutes to remove the solvent. Samples were kept at 360 K in high vacuum for half an hour to eliminate residual methanol and moisture before the measurements.

**Measurements**

A commercial Quantum Design physical property measurement system was used for the control of temperature and magnetic field. The longitudinal resistance and transverse (Hall) resistance were measured simultaneously using lock-in amplifiers (Stanford Research 830) in a six-probe Hall bar geometry. The gate voltage was supplied by a Keithley 2400 source meter. Gate voltage was swept at 330 K and a rate of of 1 mV s$^{-1}$ in high vacuum. Both the temperature and the gate voltage were held for at least 15 minutes before cooling down to ensure a homogeneous modulating throughout the sample flake. We chose a slow cooling rate (3K/min for $T$ > 70 K, 1K/min below $T$ = 70 K) to avoid possible damage of samples caused by drastically changed tension within the frozen electrolyte.

**Acknowledgements**

We would like to thank Yuanbo Zhang for discussion on ionic liquid gating, and thank Liangjian Zou and Qingming Zhang for discussion on Raman scattering. This work is supported by National Natural Science Foundation of China (NSFC), the "Strategic Priority Research Program (B)" of the Chinese Academy of Sciences, the National Basic Research Program of China (973 Program). Partial work was done in USTC center for micro- and nanoscale research and fabrication @USTC, Hefei, Anhui.

**Author contributions**

B.L. and Z.J.X. contribute equally to this work. B.L. and Z.J.X. performed the resistivity, Hall coefficient, X-ray diffraction, Raman spectrum measurements with the help of J.R.C, C.S, X.G.L, X.F.L. and N.Z.W. grew the single crystals, B.L., Z.J.X., T.W., Z.S. and X.H.C. analyzed the data. X.H.C. Z.J.X. and Z.S. wrote the paper. X.H.C. conceived and coordinated the project, and is responsible for the infrastructure and project direction. All authors discussed the results and commented on the manuscript.

**Additional information**

The authors declare no competing financial interests. Correspondence and requests for materials should be addressed to chenxh@ustc.edu.cn.

**Figure 1. Gate voltage modulation in (Li,Fe)OHFeSe based iFET. (a):** A schematic diagram of the $(Li_{0.8}Fe_{0.2})OHFeSe$ thin flake iFET device and the Hall bar contact configuration. A mixture of $LiClO_4$ and polyethylene oxide ($LiClO_4$/PEO) was employed as solid electrolyte, covering the sample and Au gate electrodes. **(b):** Gate voltage dependence of the resistance for a (Li,Fe)OHFeSe crystal flake (with a thickness of ~200 nm) during a continuous up-sweep at 330 K. The $V_g$ sweep rate is 1 mV s$^{-1}$. The inset shows an optical image of the iFET device used in our measurements.

**Figure 2. Temperature-dependent resistance of a (Li,Fe)OHFeSe iFET at various gate voltages.** The longitudinal resistance was measured from 2 K to 330 K at different gate biases ranging from 0 to 4.5 V. As the gate voltage was ramped from 0 V to 3.9 V, the mid-point critical temperature $T_c^{mid}$ gradually enhanced from 24.4 K to the highest value 43.4 K. Further increasing of the gate voltage can abruptly eliminate the superconducting state and simultaneously tune R-T curve into an insulating behavior.

**Figure 3. Hall effect measurements in (Li,Fe)OHFeSe iFET. (a):** Temperature dependence of Hall coefficient $R_H$ at different gate voltages. The thickness ($t$) of the measured (Li,Fe)OHFeSe flake is 210 nm. $R_H = (R_{xy}/B)t$ was calculated by linear fitting of $R_{xy}$ versus B plot from -5 T to 5 T in the superconducting phase, and linear fitting of $R_{xy}$ from 2.5 T to 9 T after subtracting the zero-field value in the insulating phase (see the Supplementary Information). Error bars represent the uncertainty of linear fitting. (**b**): The lower and upper panels show the Hall coefficient $R_\mathrm{H}$ and Hall number $n_\mathrm{H}= 1/R_\mathrm{H} e$ as a function of $V_\mathrm{g}$ at 70 K, respectively. Hall number per Fe site in the FeSe layer was calculated based on the unit cell volume reported in Ref.10. The filled and open circles correspond to the electron-type and hole-type carriers, respectively.

**Figure 4. Phase diagram of electronic phases as a function of gate voltage $V_\mathrm{g}$ for (Li,Fe)OHFeSe**. The boundary of SC phase and metal phase is determined by the mid-point critical temperature $T_\mathrm{c}^{mid}$. Error bars were defined as the width of SC transition. Beyond the optimal $T_\mathrm{c}$ the system immediately enters an insulator phase with increasing $V_\mathrm{g}$ and shows a sharp phase boundary between the SC/metallic phases and the insulating regime.

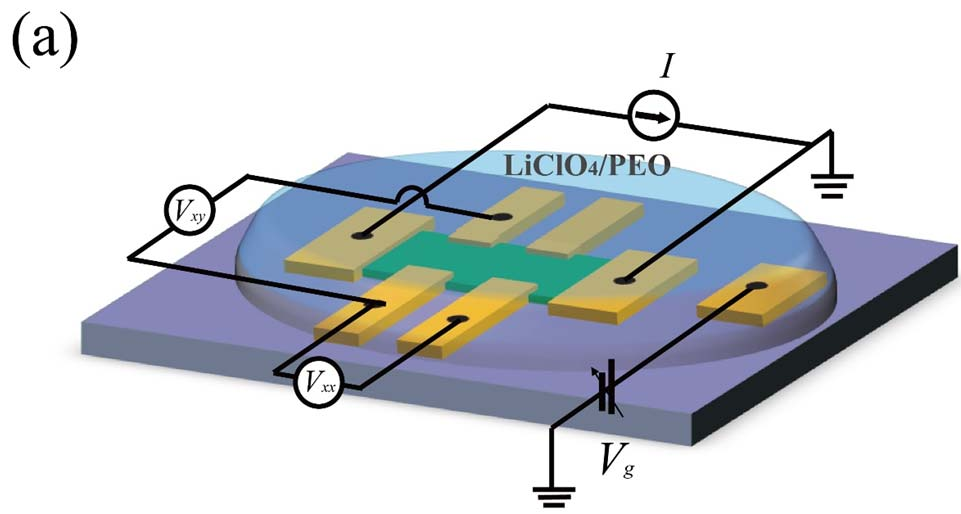

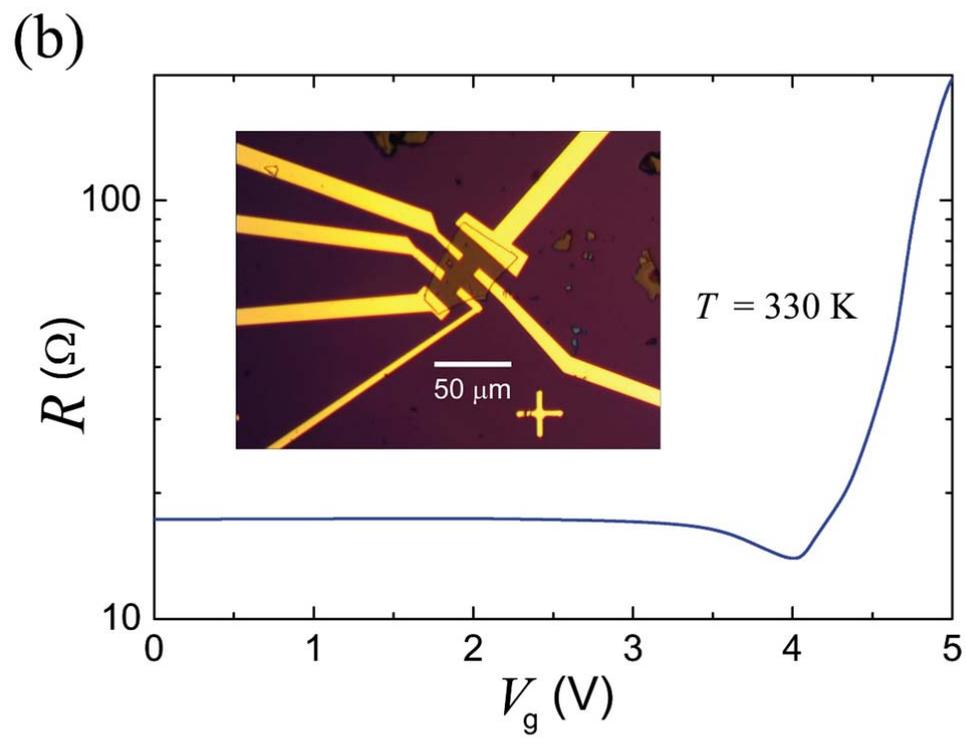

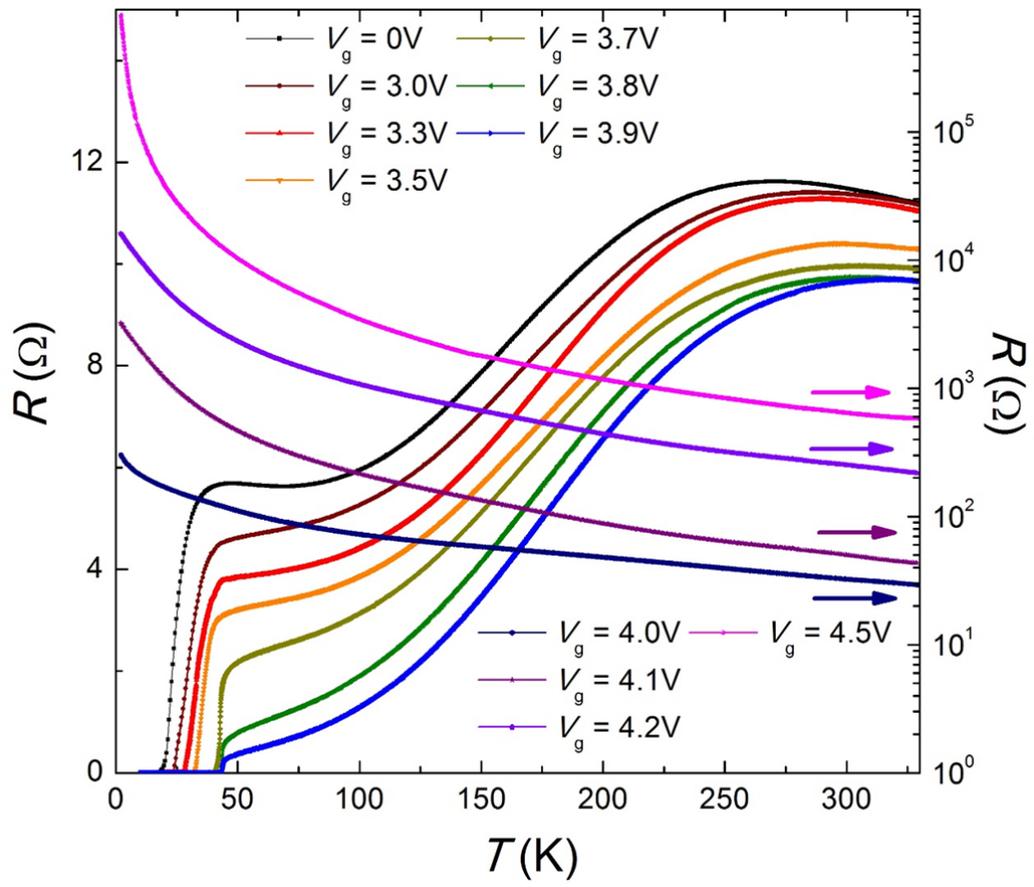

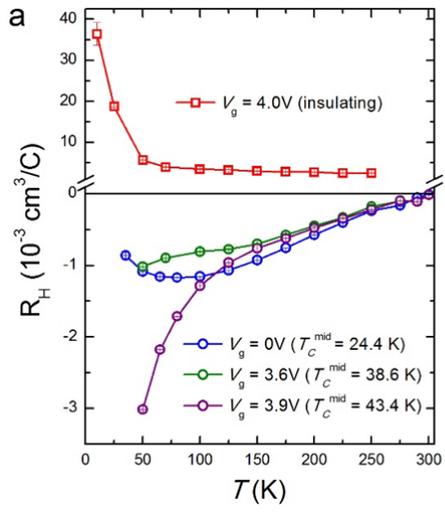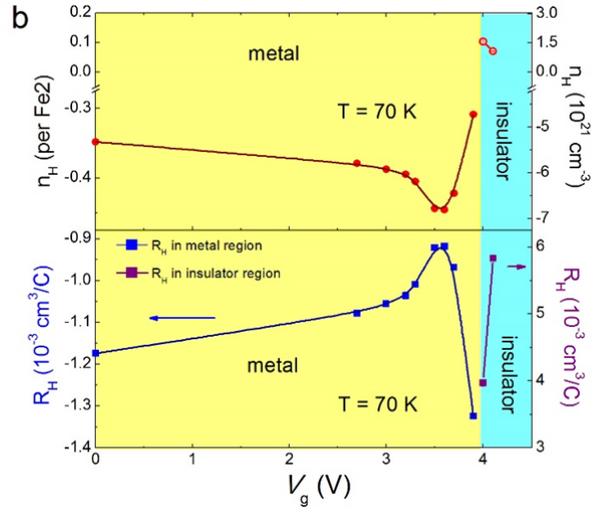

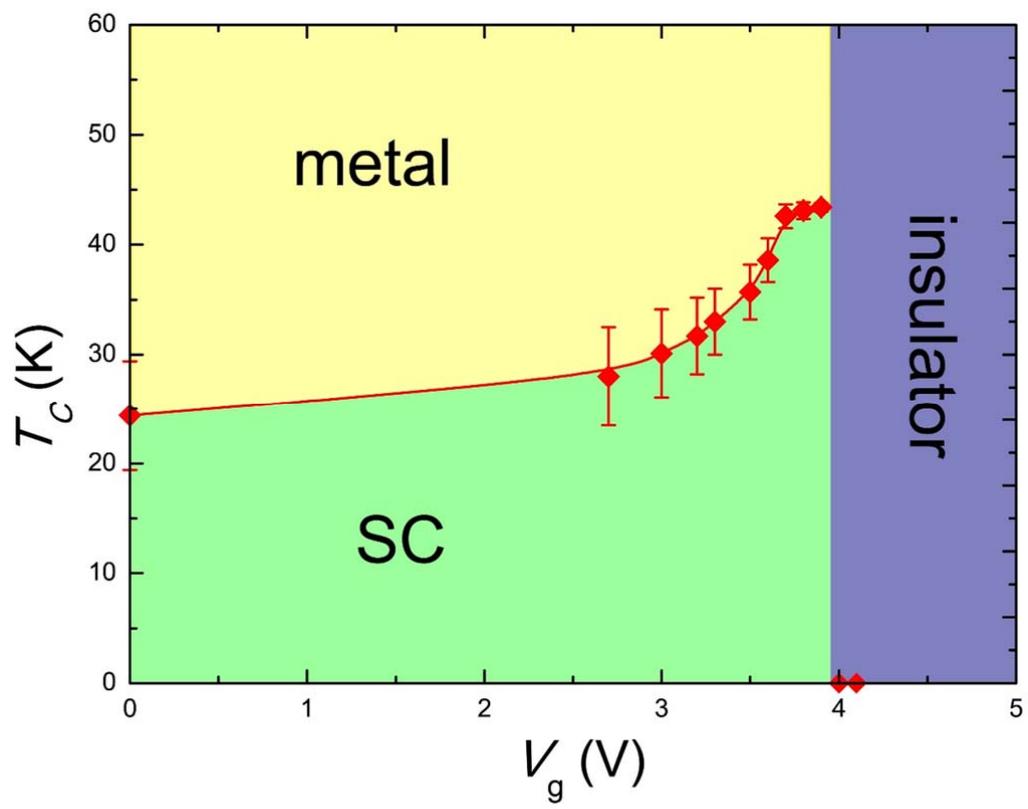